# Two-dimensional non-volatile valley spin valve

Kai Huang,[1] Kartik Samanta,[1] Ding-Fu Shao,[2] and Evgeny Y. Tsymbal[1,*]

[1] *Department of Physics and Astronomy & Nebraska Center for Materials and Nanoscience, University of Nebraska, Lincoln, Nebraska 68588-0299, USA*

[2] *Key Laboratory of Materials Physics, Institute of Solid-State Physics, HFIPS, Chinese Academy of Sciences, Hefei 230031, China*

**Abstract:** A spin valve represents a well-established device concept in magnetic memory technologies, whose functionality is determined by electron transmission being controlled by the relative alignment of magnetic moments of the two ferromagnetic layers. Recently, the advent of valleytronics has conceptualized a valley spin valve (VSV) – a device that utilizes the valley degree of freedom and spin-valley locking to achieve a similar valve effect without relying on magnetism. In this study, we propose a non-volatile VSV (n-VSV) based on a two-dimensional (2D) ferroelectric semiconductor where the resistance of the n-VSV is controlled by the ferroelectric domain wall between the two uniformly polarized domains. Focusing on the 1T'' phase of $MoS_2$, which is known to be ferroelectric down to a monolayer and using density functional theory (DFT) combined with the quantum-transport calculations, we demonstrate that switching between the uniformly polarized state and the state with oppositely polarized domains separated by a domain wall results in resistance change of as high as $10^7$. This giant VSV effect occurs due to transmission being strongly dependent on matching (mismatching) the valley-dependent spin polarizations in the two domains with the same (opposite) ferroelectric polarization orientations, when the chemical potential of 1T''-$MoS_2$ lies within the spin-split valleys. Our work paves a new route for realizing high-performance nonvolatile valleytronics.

## Introduction

A spin valve is a spintronic device widely used for information technologies. [1-3] It typically consists of two ferromagnetic layers separated by a thin non-magnetic spacer layer, either metallic or insulating, in the latter case being known as a magnetic tunnel junction. The functional property of a spin valve is a change between high and low transmission states depending on whether the magnetization directions of the two magnetic layers are parallel or antiparallel. In the parallel state, the spin configurations of the magnetic electrodes align, leading to high transmission, while in the antiparallel state, their misalignment leads to low transmission. [4,5] This results in a giant (tunneling) magnetoresistance effect useful for information readout in modern hard-drive read heads and magnetic random-access memories. [6,7]

Recently, with the advancement of two-dimensional (2D) van der Waals (vdW) materials, the concept of a valley spin valve (VSV) has emerged. [8-10] This idea utilizes a valley degree freedom to modulate the spin configuration at the Fermi surface and create a spin valve. Valleys are separated energy extrema in the electronic band structure in the momentum space. [11-20] In a nonmagnetic material with space inversion ($\hat{P}$) and time reversal ($\hat{T}$) symmetries, the valleys at non-time-reversal-invariant momenta (non-TRIMs), $+K$ and $-K$, are spin degenerate due to a combined $\hat{P}\hat{T}$ symmetry. Application of an external electric field $E$ breaks $\hat{P}$ symmetry, leading to spin splitting with opposite spin polarizations at valleys $+K$ and $-K$ due to $\hat{T}$ symmetry. Changing the sign of $E$ is equivalent to the $\hat{P}$-symmetry operation, under which the momentum $+K$ transforms to $-K$ while the spin remains invariant. The valley-dependent spin polarization is thus fully reversed by an electric field of opposite sign. This property makes valley materials useful to build an VSV. Using two independent gates to control the spin polarization of the valleys in the two separated regions of the valley material under the applied electric fields allows the control of transmission across the gates. The transmission of the VSV is high or low depending on the electric fields at the gates pointing in the same or opposite directions, respectively. So far, 2D materials, such as stanine, germanene, [9] and $(Pt/Pd)_2HgSe_3$ [10], have been predicted to develop a valley spin polarization under an applied electric field and thus to be able to constitute a VSV. In addition to memory elements, VSVs can serve as logic gates that satisfy the concatenation requirement, where the output voltage of the logic gate can be used as an input of a successive gate [21].

While the proposed concept of an VSV is interesting from the fundamental point of view and useful for practical applications, it suffers from the inability of non-volatile performance. Non-volatile memories, as opposed to volatile, allow retaining data even when not powered. The proposed VSV, however, requires an external electric field to retain the data and thus is volatile. It would be desirable from the practical perspective to develop a concept of a non-volatile VSV (n-VSV) which can potentially serve as the key element of low-power non-volatile memories and logic. At the same time, from the fundamental science perspective, a new knowledge about the properties of 2D ferroic materials exhibiting valley spin polarization is required to develop an n-VSV.

2D vdW ferroelectrics with their spontaneous electric polarization are good candidates to realize an n-VSV. Since the polarization naturally breaks $\hat{P}$ symmetry, a finite spin splitting is expected at valleys $+K$ and $-K$ even in the absence of an electric field. (Fig. 1(a)) If such a 2D ferroelectric is in a single domain state (i.e. polarization $P$ is uniform across the structure) and the Fermi level lies within the valleys, the spins states at each valley ($+K$ or $-K$) match along the transport direction and electron transmission is high (Fig. 1(a)). On the contrary, if a 2D ferroelectric has two domains with antiparallel polarization $P\uparrow$ and $P\downarrow$, the spin sates at each valleys mismatch along the transport direction across the two domains, leading to a low transmission. (Fig. 1(b))

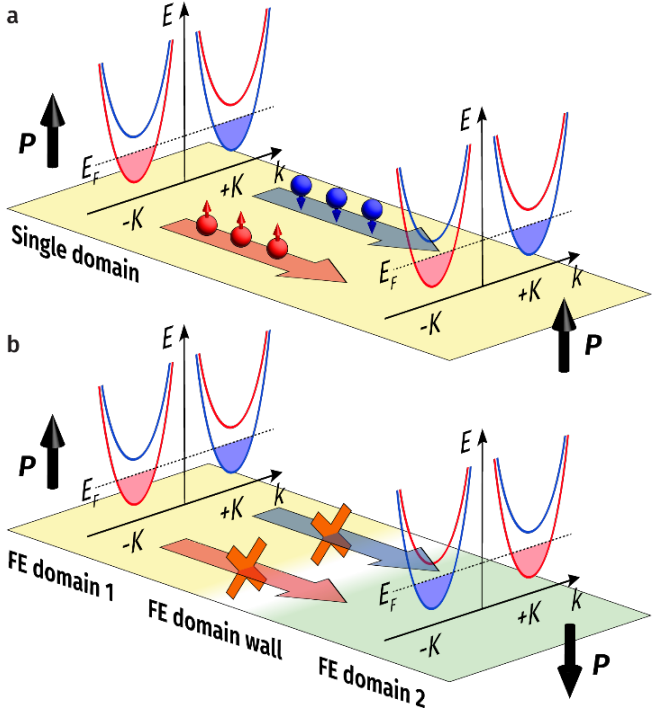

**FIG. 1**: Schematic of n-VSV based on a 2D ferroelectric material. (**a**) Spin-split band structures near $\pm K$ valleys for a single domain ferroelectric ($P\uparrow$). At each valley ($+K$ or $-K$), spin states at the Fermi energy ($E_F$) match along the transport direction and electron transmission is high. Elections with spin up (down) are denoted by red (blue). (**b**) Spin-split band structures near $\pm K$ valleys for a ferroelectric representing two opposite ferroelectric domains ($P\uparrow$ and $P\downarrow$) separated by a domain wall. At each valley ($+K$ or $-K$), spin states at $E_F$ mismatch along the transport direction and electron transmission is low.

Among the 2D materials, 1H or 2H $MoS_2$ are widely used to investigate valley-dependent effects. [22-25] However, their point-group symmetries are inconsistent with ferroelectricity. The same is true for 1T or 1T' phases of $MoS_2$ [26,27]. Recently, a ferroelectric 1T'' phase has been confirmed experimentally to occur in $MoS_2$ grown under appropriate conditions [28-32]. The 1T'' phase preserves ferroelectricity down to a monolayer limit and hosts valleys at $\pm K$ points [29,30]. These properties make 1T'' $MoS_2$ a good candidate to realize an n-VSV.

In this work, based on density functional theory (DFT) and quantum transport calculations, we predict that switching of monolayer 1T'' $MoS_2$ between the uniformly polarized state (*UP state*) and the state of two domains with opposite polarization separated by a domain wall (*DW state*) results in a giant change of transmission up to a factor of $10^7$. We show that the effect arises due to the valley spin polarization in monolayer 1T'' $MoS_2$ changing sign with reversal of electric polarization leading to spin mismatch when the UP state is flipped to the DW state.

**Monolayer 1T'' $MoS_2$**

1T'' phase $MoS_2$ is a distorted variant of the centrosymmetric 2D 1T phase $MoS_2$. [33-37] In its undistorted form, 1T $MoS_2$ consists of a trilayer structure with a molybdenum (Mo) atom surrounded by an octahedron of six sulfur (S) atoms. This undistorted 1T phase $MoS_2$ is highly unstable as a monolayer [38] and typically transforms into other more stable phases. [39-42] Among these, the 1T'' $MoS_2$ [28,32,43-46] which exhibits both ferroelectricity and valleys at non-TRIM points, [29] is particularly suitable for an n-VSV.

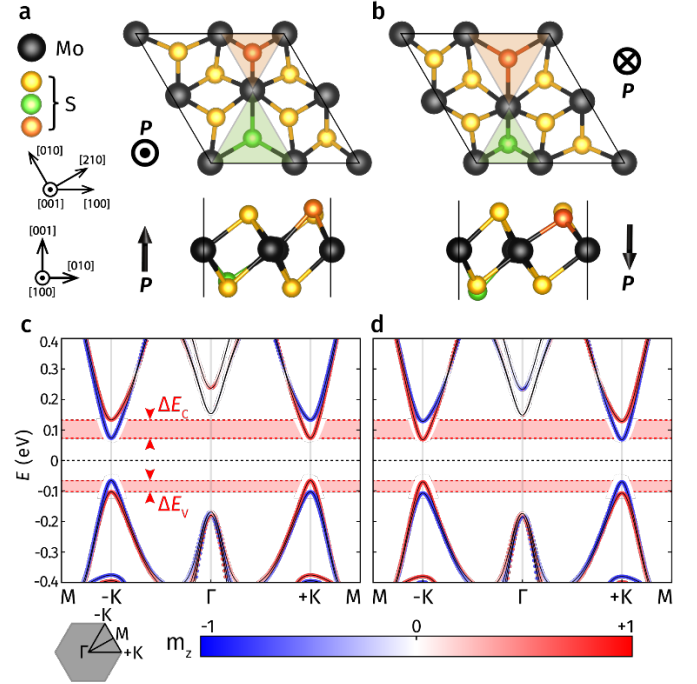

**FIG. 2**: (**a,b**) Structure of 1T'' $MoS_2$ unit cell in the top view (top) and side view (bottom) for the $P\uparrow$ state (a) and $P\downarrow$ state (b). The orange and green triangles denote Mo trimers. (**c,d**) Band structures for $P\uparrow$ state (c) and $P\downarrow$ state (d) 1T'' $MoS_2$. Spin projection to the direction of magnetization $m_z$ is shown in color. The horizontal dashed line indicates the Fermi energy ($E_F$). Hexagonal Brillouin zone and high symmetry points are shown at bottom left.

The unit cell of 1T'' $MoS_2$ contains 4 Mo atoms and 8 S atoms and belongs to the polar space group $P3m1$, indicating its ferroelectric nature. The cell is 2×2 times as large as the centrosymmetric 1T phase unit cell, due to the Mo atoms forming trimers. (Fig. 2(a,b)) In the $P\uparrow$ state, as shown in Fig. 2(a), the green Mo trimer has longer Mo-Mo distances while the orange trimer has shorter distances. This distortion pulls the green S atom closer to the middle Mo plane and pushes the orange S atom away. Other S atoms are also affected, resulting in overall S anions movement in the $-z$ direction and polarization in the $+z$ direction ($P\uparrow$). In the $P\downarrow$ state, it can be achieved by the $\hat{P}$ operation of $P\uparrow$, these movements are reversed, leading to polarization in the $-z$ direction.

Our DFT calculations, as detailed in the Supporting Information, confirm the structure and polarization properties of 1T'' $MoS_2$. The in-plane lattice constant of 1T'' $MoS_2$ is 6.34 Å which is close to the values 6.40~6.44 Å in previous reports. [28,43] The polarization (dipole moment per area) of the 2D material is calculated to be 0.25 pC/m along $z$-axis, aligning closely the value 0.24 pC/m with past studies. [28] The polarization is also confirmed by the non-zero $z$ displacement of the centers of positive and negative ions, $\bar{d}_z^S - \bar{d}_z^{Mo} = -0.0052$Å for the $P\uparrow$ state, where $\bar{d}_z^S$ is the averaged $z$ displacement of 8 S atoms and $\bar{d}_z^{Mo}$ is the averaged $z$ displacement of 4 Mo atoms in the unit cell.

Fig. 2(c,d) shows the band structure of 1T'' $MoS_2$ in $P\uparrow$ and $P\downarrow$ states calculated with inclusion of SOC. The material exhibits a band gap of 0.14 eV, consistent with previously reported values ranging from 0.1 to 0.19 eV. [28,29,43] We find valleys of conduction band minimum (CBM) and valence band maximum (VBM) at the $\pm K$ points, where the Zeeman-like spin splitting for spin-up and spin-down electrons is induced by $\hat{P}$ symmetry breaking. The spin splitting is 60 meV for the valleys above the Fermi level and 37 meV for the valleys below it. We denote the energy windows for the valley-dependent spin splitting as $\Delta E_C$ and $\Delta E_V$ for CBM and VBM, respectively. Due to $\hat{T}$ symmetry, the spin splittings at the $+K$ and $-K$ points are opposite. The valley-dependent spin polarizations of the $P\uparrow$ and $P\downarrow$ states are also opposite, as they are related by $\hat{P}$ symmetry.

**Atomic and band structure of n-VSV**

An n-VSV requires switching between UP and DW states. To model this, we combine the orthorhombic supercells of $P\uparrow$ and $P\downarrow$ states 1T'' $MoS_2$ with lattice vectors of these supercells were oriented along the [210] and [010] directions, respectively. As denoted in Fig. 3(a), the orthorhombic supercells are marked by blue dashed boxes. For the calculation, three $P\uparrow$ orthorhombic supercells and three $P\downarrow$ orthorhombic supercells are combined. During the structural relaxation process, the regions numbered 5 through 8 and 17 through 20 (Fig. 3(b)) were kept fixed in the middle of both $P\uparrow$ and $P\downarrow$ domains, while other areas were allowed to relax to identify stable DW structures. This process reveals two possible DWs, designated by DW1 and DW2. Using the polarization displacement $\bar{d}_z^S - \bar{d}_z^{Mo}$ shown in Fig. 3(b), we observed that the displacement increases abruptly across the DWs.

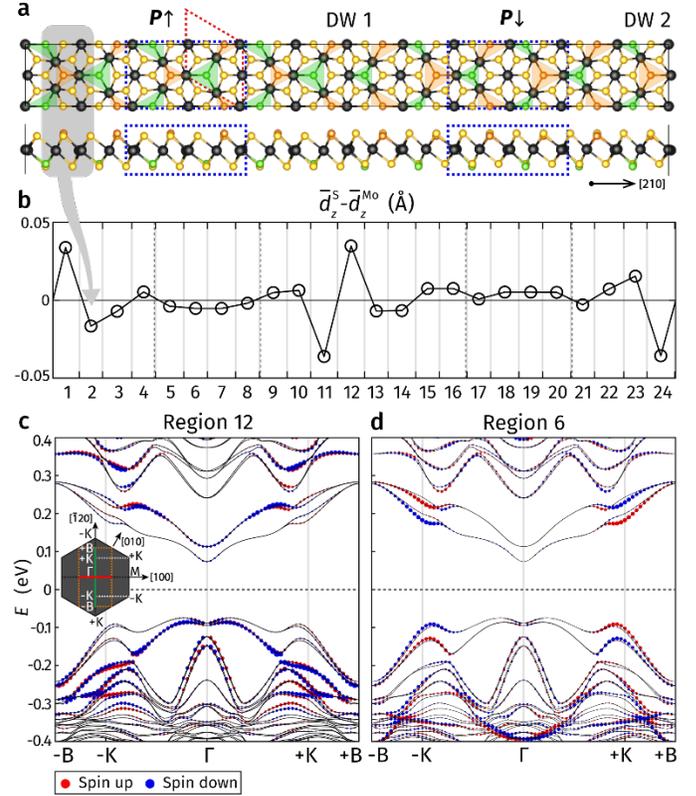

**FIG. 3**: (**a**) Top view (top) and side view (bottom) of the structure with DW separating $P\uparrow$ and $P\downarrow$ domains of 1T'' $MoS_2$. Atoms within the blue boxes represent fixed pristine domains while other atoms are subjected to structural relaxation. The red dashed rhombus indicates the unit cell of the $P\uparrow$ domain. The orange and green triangles denote Mo trimers. (**b**) $\bar{d}_z^S - \bar{d}_z^{Mo}$ value for atoms in different regions in the structure of structurally relaxed DWs, as region 2 shown in grey as an example, numbered from 1 to 24, in the supercell structure. (**c,d**) Band structures (black lines) of the domain wall structure of 1T'' $MoS_2$. Spin projection to the direction of magnetization $m_z$ is shown in color. The width of the fat band represents the electron contribution from the atoms in region 12 (c), 6 (d). Insert: Brillouin zone of the 1T'' $MoS_2$ hexagonal unit cell (dark grey), orthogonal super cell (orange), and the DW structure along [100] (green) and [$\bar{1}$20] (red). Here directions are in reciprocal coordinates according to the hexagonal unit cell.

Fig. 3(c,d) shows the electronic band structure of the relaxed structure illustrated in Fig. 3(a). Due to the shape of the supercell, the hexagonal Brillouin zone of the primitive cell is folded to the line along the [$\bar{1}$20] direction, perpendicular to transmission (green line in the insert of Fig. 3(c)). In contrast to the bulk band structure shown in Fig. 2(c), the CBM and VBM in the DW supercell emerge at the $\Gamma$ point. We project the regions shown in Fig. 3(a) to the band structure and find that the band extrema at the $\Gamma$ point is majorly contributed by region 12 associated with the DW1 (Fig. 3(c)) and region 24 associated with the DW2 (Fig. S2(a)), indicating that the distortion at the DW significantly influences the local electronic structure. On the other hand, the regions 6 (Fig. 3(d)) and 19 (Fig. S2(b)) at the center of the domains contributed dominantly to the energy

local extrema at $\pm K$. The spin splitting above Fermi level at $\pm K$ is 47 meV, and 33 meV below Fermi level, indicating that the valleys are well maintained away from the DWs.

**Conductance and ON/OFF ratio of n-VSV**

We then construct an n-VSV structure assuming that the transport direction is along the [210] axis of 1T'' MoS$_2$. For the UP states, we directly calculate transmission (and hence conductance) of a homogeneous 1T'' MoS$_2$. Fig. 4(b) shows conductance $G_{k_\parallel}(E_F)$ as functions of the Fermi energy $E_F$ and transverse wavevector $k_\parallel$ (along $[\bar{1}20]$). Qualitatively, the conductance pattern reflects the band structure with pronounced valleys at the $\pm K$ points. When $E_F$ lies at the midgap, conductance is zero due to no available electronic states for transport. As $E_F$ enters the valley near CBM, conductance becomes non-zero due to population of the lower spin band. This is evident from the sharp increase in the total conductance $G = \sum_{k_\parallel} G_{k_\parallel}$ at this energy (blue line in Fig. 4(d)). The conductance value is $G_{k_\parallel}(E_F) = 1\ e^2/h$ (orange color in Fig. 4(b)) for $E_F$ within $\Delta E_C$, due to only one spin band appearing in the valleys in this energy window. Similar behavior is observed for $E_F$ within $\Delta E_V$.

For the DW states, as shown in Fig. 4(a), we use 1T'' MoS$_2$ with $P\uparrow$ and $P\downarrow$ domains as the electrodes, and the block of DW1 containing region 5 to region 20 in Fig. 3 as the scattering region. We find that conductance is significantly affected by the presence of the DW. $G_{k_\parallel}(E_F)$ is negligibly small at $\pm K_\parallel$ points for $E_F$ within both $\Delta E_C$ and $\Delta E_V$, as seen from Fig. 4(c). This suppression of conductance occurs because the valley-dependent spin polarizations in the two domains are opposite. For the $E_F$ far away from CBM and VBM where the valley-dependent spin splitting becomes small, the total conductance $G$ increases sharply as both spin bands get populated (orange line in Fig. 4(d)) enabling scattering between bands with the mixed spin polarization.

Fig. 4(e) shows the ON/OFF ratio, $\eta = (G_{UP} - G_{DW})/G_{DW}$, where $G_{UP}$ and $G_{DW}$ are conductance for the UP (ON) and DW (OFF) states, respectively. It is seen that in the energy range of $\Delta E_C$ and $\Delta E_V$, the n-VSV exhibits a giant resistive switching effect with $\eta$ being as high as $\sim 10^7$ above the bandgap and $\sim 10^6$ below the bandgap. This huge effect reflects the significant change in conductance between UP and DW states (Fig. 4(d)) due to locking between the spin and electric polarization and represents the key property of an n-VSV.

**Discussion and summary**

We would like to emphasize that the predicted giant VSV effect is driven by SOC which produces Zeeman-like splitting of the spin bands at non-TRIM momentum valleys. This is, in particular, evident from our calculations without SOC, showing loss of the effect (Fig. S5). Therefore, apart from 1T'' MoS$_2$, 2D ferroelectrics with stronger SOC are expected to be beneficial for the design of an n-VSV.

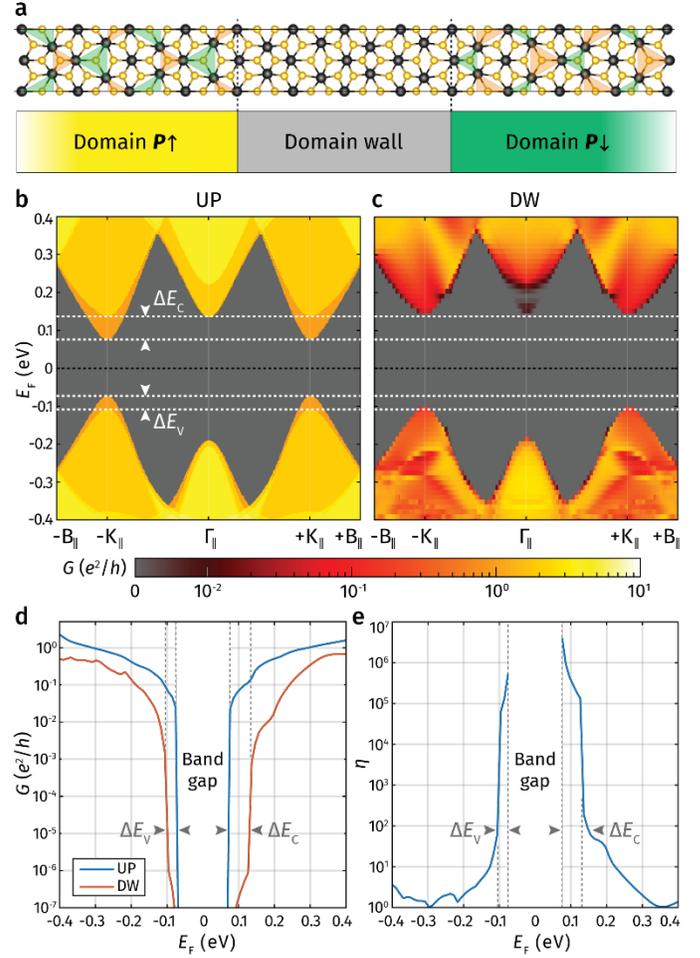

FIG. 4: (**a**) n-VSV structure used for transmission calculations. (**b,c**) Conductance of the n-VSV in UP state (b) and DW state (c) as functions of the Fermi energy $E_F$ and electron momentum perpendicular to the transmission direction ($k_\parallel$). The conductance is denoted by color as shown in the color bar in the log scale. (**d**) Total conductance as a function of $E_F$ for the UP and DW states. (**e**) ON/OFF ratio as a function of $E_F$.

The transport properties of an n-VSV are expected to depend on the DW width between the two domains with opposite polarizations. We therefore checked the stability of the DW in Fig. 3(a,b) by increasing the size of the supercell in the longitudinal direction and performing structural relaxation, as described in Supporting Information. Our calculations indicated that the shape and width of the DW remained largely unchanged, showing its structural stability. The stability of other possible DWs is discussed in Supporting Information.

As seen from Fig. 3(a,b), the DW1 and DW2 have somewhat different atomic structures. We therefore calculated the appearance of the VSV effect for a device with DW2 in the scattering region. Our results indicated qualitatively very similar behavior (Fig. S4).

We note that the ON/OFF ratio is controlled by the valley-dependent spin matching in momentum space, which is strongly influenced by the choice of transport direction. For example, the VSV effect is expected to be negligible in a 1T'' MoS$_2$-based VSV if the transport is along the [010] direction. In this case, the valleys at $\pm K$ overlap at the same $k_\parallel$, and hence momentum-dependent spin polarizations for these valleys cancel out. The ON/OFF ratio in this device is then purely due to the domain wall scattering.

The Fermi energy is the critical parameter that controls the VSV effect. Doping is required to shift the Fermi level to $\Delta E_C$ or $\Delta E_V$ for the giant VSV effect to occur. This may be achieved by adding impurities to the bulk material or by a charge transfer from the substrate or capping layer.

Overall, our results demonstrate the viability of using 1T'' MoS$_2$ as the core material for an n-VSV. By exploiting the intrinsic ferroelectric properties of this 2D vdW material, we demonstrate non-volatile control over valley-dependent spin polarization and electron transmission without the need for an external electric field. The resulting nVSV device shows a remarkably high resistance ratio between the ON and OFF states, reaching up to $10^7$, thus validating the concept of non-volatile valleytronics. This advancement paves a new route for more energy-efficient and high-performance valleytronic devices, highlighting the importance of 2D ferroelectrics in future electronic applications.

## Associated content

The Supporting Information is available free of charge at [link here]

It includes additional details of computational methods; stability of the DWs; alternative DW structures; transmission across DW2; transmission without SOC.

## Author information

**Corresponding Author**

Evgeny Y. Tsymbal – https://orcid.org/0000-0002-6728-5480; Email: tsymbal@unl.edu

**Authors**

Kai Huang – https://orcid.org/0000-0001-5527-3426; Email: kai.huang@huskers.unl.edu

Kartik Samanta – https://orcid.org/0000-0003-3496-3238; Email: ksamanta2@unl.edu

Ding-Fu Shao – https://orcid.org/0000-0002-2732-4131; Email: dfshao@issp.ac.cn

**Notes**

The authors declare no competing financial interest.

## Acknowledgements

This work was supported by the grant number DE-SC0023140 funded by the U.S. Department of Energy, Office of Science, Basic Energy Sciences (K. H., K. S., E. Y. T). Computations were performed at the University of Nebraska Holland Computing Center.